# Resolving the Magnetic Ground-State Controversy in $RuO_2$ through A Flat Magnetic Energy Landscape

Tianxiao Liang[1], Fanhan Kong[1], Jijun Zhao[2,3], and Xue Jiang[2,3*]

[1]Key Laboratory of Material Modification by Laser, Ion and Electron Beams (Dalian University of Technology), Ministry of Education, Dalian 116024, China

[2]Guangdong Basic Research Center of Excellence for Structure and Fundamental Interactions of Matter, Guangdong Provincial Key Laboratory of Quantum Engineering and Quantum Materials, School of Physics, South China Normal University, Guangzhou 510006, China

[3]Quantum Science Center of Guangdong-Hong Kong-Macao Greater Bay Area, Shenzhen-Hong Kong International Science and Technology Park, Shenzhen 518000, China

## Abstract

Rutile $RuO_2$ is a prominent candidate for altermagnetism, yet its magnetic ground state remains highly controversial, with experiments reporting either a nonmagnetic state or altermagnetic order. Here, we develop a generalized environment-dependent spin-lattice framework that unifies localized Heisenberg exchange, itinerant Stoner magnetism via Landau spin fluctuations, and spin-orbit-coupling-mediated spin-lattice interactions. Parameterized from a high-throughput first-principles database using machine-learning and solved by large-scale Monte Carlo simulations, the framework reveals an exceptionally flat magnetic energy landscape in $RuO_2$, where the nonmagnetic state lies nearly degenerate with multiple altermagnetic configurations. We find that material perturbations, exemplified by intrinsic defects, select distinct magnetic ground states primarily by modifying the localized Heisenberg exchange, with perturbation-induced itinerant Stoner polarization provides an essential secondary contribution. Spin-orbit coupling controls the orientation and stability of the Néel vector, but does not determine the emergence of long-range magnetic order. These results provide a unified explanation for the conflicting experimental observations and establish a general microscopic framework for understanding how material perturbations select competing magnetic ground states in systems with nearly flat magnetic energy landscapes.

*Introduction*—The recent discovery of altermagnetism (AM), a distinct class of collinear magnetic order beyond conventional ferromagnetism (FM) and antiferromagnetism (AFM) [1-3], has introduced a new paradigm for spin-dependent phenomena. Altermagnets retain the zero net magnetization of AFMs while exhibiting momentum-dependent spin splitting, enabling unconventional responses such as the anomalous Hall effect (AHE), spin-splitting torque, and novel spin-dependent transport phenomena without stray fields [4-7]. Among the proposed AM materials, rutile $RuO_2$ has emerged as one of the most extensively investigated systems. Establishing its magnetic ground state is therefore essential for elucidating the origin of diverse magnetic phenomena reported in $RuO_2$. However, despite extensive theoretical and experimental efforts, the magnetic ground state of $RuO_2$ remains unresolved, with experiments reporting contrasting signatures of magnetic order and nonmagnetism [8].

A wide range of experiments, including polarized neutron diffraction (PND) [9], angle-resolved photoemission spectroscopy (ARPES) [10], anomalous Hall effect (AHE) [4, 11], spin-splitting torque (SST) [12-15], and inverse altermagnetic spin-splitting effect (IASSE) measurements [16, 17], have been interpreted as evidence for an altermagnetic ground state in $RuO_2$. These observations are further supported by spin-group symmetry analyses, which identify $RuO_2$ as a candidate platform for altermagnetic spin splitting [1, 18], the crystal Hall effect [4, 18] and anomalous crystal thermal transport [19]. In contrast, muon spin rotation and relaxation (μSR) [20, 21], high-resolution ARPES [22], x-ray diffraction (XRD) [23], electrical and thermal transport measurements [24, 25], and optical spectroscopy [26, 27] have reported no evidence of bulk magnetic order, pointing instead to a nonmagnetic ground state. Moreover, the absence of IASSE signatures in some $RuO_2$ thin films further indicates that altermagnetic responses may not be universal across samples [28, 29]. Recent theoretical work by Smolyanyuk *et al.* [30] demonstrated that Ru vacancies can induce magnetic instability at realistic Hubbard $U_{eff}$ values and strongly favor altermagnetic order, suggesting that subtle sample-dependent perturbations may play a decisive role in determining the magnetic state. However, why $RuO_2$ is so is so susceptible to such perturbations, and how they select among competing magnetic states, remains unclear. A quantitative microscopic framework connecting sample-dependent perturbations to the experimentally observed magnetic behavior under realistic conditions is still lacking.

Here, we identify an exceptionally flat magnetic energy landscape as the microscopic origin of the long-standing magnetic-state controversy in $RuO_2$, in which the nonmagnetic state is nearly degenerate with multiple altermagnetic states. This near degeneracy renders the magnetic state highly susceptible to material perturbations, allowing subtle variations in sample conditions to select distinct ground states. Using intrinsic defects as a representative perturbation, we develop a generalized environment-dependent spin-lattice Hamiltonian that incorporates both localized Heisenberg exchange and itinerant Stoner magnetism. Parameterized from a high-throughput DFT database using machine-learning techniques, the model is solved by large-scale MC simulations. We find that (i) materials perturbations, represented here by defect concentrations, predominantly select the magnetic state through modifications of the Heisenberg exchange with perturbation-induced Stoner polarization providing a secondary yet non-negligible contribution; (ii) spin-orbit coupling governs the orientation of Néel vector but does not determine the emergence of long-range magnetic order. Together, these results provide a unified microscopic explanation for the disparate experimental observations. Beyond resolving the magnetic state controversy in $RuO_2$, our work establishes a general mechanism by which material perturbations select among competing low-energy magnetic states on nearly flat energy landscapes. More broadly, this perturbation-induced state-selection mechanism provides a general route to controlling competing magnetic phases in correlated materials with nearly flat energy landscapes.

*Results and discussion*—We develop a generalized environment-dependent spin-lattice Hamiltonian that incorporates the essential magnetic interactions relevant to correlated oxides while explicitly accounting for the local atomic environment. The Hamiltonian is given by **Eq. 1**,

$$H = -\frac{1}{2}\sum_{i,j} J_{ij}\boldsymbol{S}_i \cdot \boldsymbol{S}_j + \sum_i [A_i(\boldsymbol{R})|\boldsymbol{S}_i|^2 + B_i(\boldsymbol{R})|\boldsymbol{S}_i|^4] + \sum_i \boldsymbol{K}_i(\boldsymbol{R}) \cdot \boldsymbol{S}_i + \sum_i \boldsymbol{S}_i^T \cdot \boldsymbol{\Lambda}_i(\boldsymbol{R}) \cdot \boldsymbol{S}_i + \sum_{ij} \boldsymbol{S}_i^T \boldsymbol{J}_{ij}^{ani}(\boldsymbol{R})\boldsymbol{S}_j + E_0 \quad (1)$$

where the first term describes the isotropic Heisenberg exchange interaction [31], the second term is an onsite Landau expansion that captures longitudinal fluctuations of the local magnetic moments [32], the third and fourth terms describe spin-lattice couplings mediated by SOC [33], and the fifth term accounts for anisotropic exchange interactions, including the Kitaev interaction [34] and Dzyaloshinskii-Moriya interactions (DMI) interactions [35]. The physical origin and detailed derivation of each term are provided in the Supplemental Material (SM) [36].

Unlike conventional spin Hamiltonians with fixed interaction parameters, the model parameters in **Eq. 1** explicitly depend on the local atomic configuration $\boldsymbol{R}$, allowing the competing magnetic interactions to evolve continuously with the local materials environment. This environment dependence provides a unified microscopic description of how intrinsic defects, lattice distortions, carrier doping and other sample-dependent perturbations renormalize exchange interaction, itinerant Stoner polarization, spin-lattice coupling, and magnetic anisotropy within a unified microscopic framework. In particular, the onsite Landau coefficient $A_i(\boldsymbol{R})$ captures variations of the local Stoner instability induced by the surrounding atomic environment, providing an effective description of the itinerant magnetic channel. The resulting Hamiltonian therefore unifies localized Heisenberg exchange and itinerant Stoner magnetism within a single environment-dependent framework.

Among these terms, the spin-lattice coupling tensors $\boldsymbol{K}_i$ and $\boldsymbol{\Lambda}_i$ are defined as,

$$\boldsymbol{K}_i = \nabla_{\boldsymbol{R}}\, \rho_i \quad (2)$$

$$\boldsymbol{\Lambda}_i^{jk} = \frac{\partial^2 \rho_i}{\partial \boldsymbol{R}_j\, \partial \boldsymbol{R}_k} \quad (3)$$

here $\rho_i$ represents the local electronic charge distribution surrounding atom $i$. The Hamiltonian parameters are obtained by machine-learning assisted fitting to a high-throughput density functional theory (DFT) database [36, 37]. Once parameterized, the model provides a quantitative framework for describing competing magnetic interactions under realistic material conditions and serves as the microscopic foundation for the large-scale Monte Carlo (MC) simulations presented below. For collinear antiferromagnets with two magnetic sublattices ($\boldsymbol{S}_1 = -\boldsymbol{S}_2 = \boldsymbol{S}$), including altermagnetic systems, the framework provides a quantitative description of magnetic ground states and finite-temperature magnetic properties. More broadly, its environment-dependent formulation makes it applicable to other two sublattice collinear magnetic systems in which local material perturbations modify competing magnetic interactions, including $RuO_2$, CrSb, or $MnO_2$.

As illustrated in **Fig. 1(a)**, rutile $RuO_2$ contains two symmetry-related Ru sublattices (Ru1 and Ru2), which form the two magnetic sublattices considered in our model. We constructed a high-throughput DFT database consisting of 215 magnetic configurations, including nonmagnetic, ferromagnetic, altermagnetic, and noncollinear states with systematically varied magnetic moments and Néel-vector orientations (**Table S1** in the SM [36]). The resulting Hamiltonian accurately reproduces the DFT energies, with a mean absolute error of 14 meV/atom over an energy range exceeding 35 eV [**Fig. 1(b)**]. Using the fitted Hamiltonian, we map the magnetic energy landscape by continuously varying the magnetic moments on the two Ru sublattices. Remarkably, the resulting landscape [**Fig. 1(c)**] exhibits an extended low-energy basin in which the NM state and multiple AM configurations are nearly degenerate. This nearly flat landscape provides a natural energetic explanation for the diversity of magnetic states observed experimentally in $RuO_2$. For comparison, analogous calculations for the experimentally established altermagnet CrSb (**Fig. S2** in the SM [36]) reveal a well-defined magnetic minimum rather than an extended low-energy basin. The exceptionally flat landscape of $RuO_2$ therefore renders its magnetic state highly susceptible to weak material perturbations, providing the energetic basis for perturbation-induced selection among competing magnetic states. Computational details, fitting procedures, and optimized parameters are provided in the SM [36].

Machine-learning-assisted fitting reveals that the Heisenberg and Landau terms dominate the magnetic energetics of pristine $RuO_2$, whereas the spin-lattice coupling and anisotropic exchange terms contribute negligible. The magnetic energy landscape it therefore well described by an effective Heisenberg-Landau Hamiltonian, indicating that SOC plays only a minor influence in determine the relative energies of competing magnetic states. Consistent with this picture, self-consistent DFT calculations show that simultaneously rotation of the two Néel vectors produces only negligible changes in the converged electronic structure, which the Ru-*d* density of states exhibits substantial overlap among different magnetic configurations [**Fig. 1(d)**]. These results demonstrate that SOC does not lift the near-degeneracy established by the Heisenberg-Landau interactions and therefore does not determine the existence of magnetic order in pristine $RuO_2$, although it can still influence the orientation of the Néel vector.

Despite its accuracy for pristine $RuO_2$, the framework has two important limitations when extended to realistic sample conditions. First, it does not explicitly capture perturbation-enhanced SOC effects that may arise from strong defects, lattice strain, or external magnetic fields. Second, magnetic parameters extracted from static DFT calculations within the mean-field approximation tend to overestimate the Néel temperature, as discussed in **Sec. 2.2** of the SM [36, 37]. We therefore construct an effective spin-lattice Hamiltonian for experimentally relevant conditions, which is given by **Eq. 4**,

$$\begin{aligned}\mathcal{H} = \quad & -\frac{1}{2}J_{\mathrm{eff}}\sum_{i,j}\mathbf{S}_i\cdot\mathbf{S}_j+\sum_i\left(A_i(\boldsymbol{R})|\mathbf{S}_i|^2+B_i(\boldsymbol{R})|\mathbf{S}_i|^4\right)\\ & +\frac{\lambda K_0}{z}\sum_{i,j}S_i^zS_j^z-g\mu_B\mathbf{B}\cdot\sum_i\mathbf{S}_i+E_0\end{aligned} \tag{4}$$

In this formulation, the dominant Heisenberg exchange and Landau terms are retained, with the effective exchange interaction $J_{\mathrm{eff}}$ and Landau coefficient $A_i(\boldsymbol{R})$ evolving with the local material environment. For the representative samples considered here, this environment dependence is parameterized by the Ru and O vacancy concentrations ($c_{\mathrm{Ru}}$ and $c_{\mathrm{O}}$), which capture sample-dependent renormalization of both the Heisenberg exchange and itinerant Stoner channels. External magnetic fields enter through the conventional Zeeman term, while the SOC-induced exchange anisotropy $\lambda$ is incorporated as an effective perturbation [33, 38]. The detailed derivation and parameterization of **Eq. 4** are provided in **Secs. 2.3** and **2.4** of the SM [36]. This effective Hamiltonian provides a unified microscopic description of magnetic behavior across realistic sample conditions, from high-quality single crystals to defect-rich thin films. We investigate five representative sample conditions corresponding to previous experimental studies [11, 20, 21, 23, 36, 39] using large-scale Monte Carlo simulations. These samples span the range of magnetic behaviors reported experimentally, from high-quality single crystals with no detectable magnetic order to defect-rich thin films exhibiting robust altermagnetic signatures. We first set $\lambda = 0$ to isolate the effects of the Heisenberg exchange and Stoner channels. The resulting effective Néel temperatures ($T_N$) and staggered magnetizations ($M_{\text{stag}}$) are used to characterize the magnetic behavior, while all sample parameters and computational details are provided in the SM [36].

We first validate the effective Hamiltonian by examining whether it quantitatively reproduces representative experimental observations under different sample conditions. As summarized in **Fig. 2**, the calculated Néel temperature $T_N$ and staggered magnetization $M_{\text{stag}}$ naturally separate the five representative samples into three distinct magnetic regimes. High quality single crystals studied by Hiraishi *et al.* [20] and Kiefer *et al.* [23] exhibit very low temperature of $T_N \approx$ 3-5 K and essentially vanishing staggered magnetization, consistent with the absence of detectable long range magnetic order in muon spin resonance and related measurements [20]. In contrast, defect-rich films reported by Feng *et al.* [11] and Dai *et al.* [39] exhibit robust magnetic order, with calculated $T_N \approx$ 300-400 K, consistent with the experimentally reported $T_N \approx$ 400 K [40]. Their robust altermagnetic character is further supported by the observation of the anomalous Hall effect (AHE) [11] and spin-dependent Peltier effect (SDPE) [39]. Between these two limits, the intermediate-quality sample studied by Keβler *et al*. [21] exhibits two distinct $T_N$ branches, spanning approximately 12-16 K and 50-150 K, placing it near the boundary between the weakly magnetic and robustly ordered regimes and capturing the coexistence of nonmagnetic and altermagnetic signals observed experimentally. The calculated staggered magnetization follows the same trend [**Figs. 2(b)** and **2(c)**], with only the defect-rich samples retaining a robust ordered moment at room temperature, whereas the other samples remain close to the nonmagnetic regime. The agreement across these representative experimental conditions demonstrates that the effective spin-lattice Hamiltonian captures the essential physics governing the sample-dependent magnetic behavior of $RuO_2$. Rather than reflecting mutually exclusive magnetic ground states intrinsic to pristine $RuO_2$, the diverse experimental observations can be understood as different realizations of competing magnetic states selected by sample-dependent perturbations within its exceptionally flat magnetic energy landscape.

Having established that the effective Hamiltonian reproduces the experimentally observed magnetic behavior, we next disentangle the contributions of the Heisenberg and Stoner channels to the sample-dependent magnetic response. Because both the effective exchange interaction $J_{\text{eff}}$ and the Landau coefficient $A_i(\boldsymbol{R})$ depend on the defect concentrations ($c_{\text{Ru}}$ and $c_{\text{O}}$), their individual contributions cannot be isolated directly from **Eq. 4**. To separate these effects, we introduce two independent dimensionless scaling factors, $\kappa_J$ and $\kappa_A$, which selectively scale the Heisenberg exchange and effective Stoner (Landau) channel, respectively (see **Sec. 2.5** of the SM [36]). The reference model correspond to $\kappa_J = 1$ and $\kappa_A = 1$ (**Table S8** in the SM [36]) and the magnetic response is assessed by varying each factor independently. As shown in **Fig. 3**, variation in $\kappa_J$ produces substantially larger changes in both $T_N$ and $M_{\text{stag}}$ than comparable variations in $\kappa_A$, demonstrating that the Heisenberg exchange interaction is the dominant contribution to magnetic ordering in $RuO_2$. The Landau Stoner channel nevertheless provides an essential secondary contribution, with its influence on $T_N$ being most pronounced in high-quality single-crystal samples, while its effect on $M_{\text{stag}}$ becomes appreciable primarily in samples supporting altermagnetic order and remains weak in the nonmagnetic regime. These results reveal a clear hierarchy between the two magnetic channels: the localized Heisenberg exchange primarily controls the emergence of long-range magnetic order, whereas the itinerant Stoner channel provides a secondary renormalization of the stability of competing magnetic states once exchange-driven order is established. This secondary contribution becomes significant because the competing states reside within an exceptionally flat magnetic energy landscape, where even relatively weak changes in the itinerant magnetic channel can appreciably modify their relative stability.

Finally, we examine how spin-orbit coupling modifies the magnetic phase diagram by performing

Monte Carlo (MC) simulations over a range of SOC-induced anisotropy strength $\lambda$. As shown in **Fig. 4**, varying $\lambda$ changes the sample-dependent Néel temperatures $T_N(\lambda)$ quantitatively, but leaves the overall topology of the magnetic phase diagram unchanged. The phase diagram naturally separates into three physically distinct regions. At temperatures above the largest sample-dependent $T_N(\lambda)$ among the representative samples, all systems are in the paramagnetic state. Because the simulations are restricted to a finite temperature window, the absence of a phase boundary above this range should not be interpreted as the absence of magnetic order at higher temperatures. The broad intermediate region is characterized by strong sample dependence: under otherwise comparable conditions, different material environments can support either weakly magnetic or robustly altermagnetic states. Its upper boundary is determined by the samples with the largest $T_N(\lambda)$, represented by the Feng *et al.* and Dai *et al.* samples, whereas its lower boundary is set by the weakly magnetic Hiraishi *et al.* and Kiefer *et al.* samples. The calculated staggered magnetizations are consistent with the contrasting experimental observations. The high-quality single crystals samples exhibit only weak short-range magnetic correlations, with $M_{stag}$ @300 K ≈ 0.05 $\mu_B$, consistent with the absence of detectable long-range order in μSR and neutron diffraction measurements. In contrast, the defect-rich samples retain robust altermagnetic order, with $M_{stag}$@300 K ≈ 1.5 $\mu_B$, consistent with the experimentally observed AHE and SDPE signals. The intermediate-quality samples reported by Keβler *et al.* lie close to the phase boundary, naturally accounting for the coexistence of magnetic and nonmagnetic signatures observed experimentally.

A second characteristic feature of **Fig. 4** is the narrow detection-limited regime at low temperatures ($\lambda \lesssim 0.1$ and $T \lesssim 5$ K), where the minimum sample-dependent Néel temperature remains below 5 K. Magnetic order may exist in principle within this regime, but the corresponding ordered moment remains below the sensitivity of currently experimental probes, with $M_{\rm stag}^{\rm min}$@ 300 K < $10^{-3}$ $\mu_B$. This regime is naturally associated with the high-quality single-crystal samples of Hiraishi *et al.* and Kiefer *et al.*, for which the weak exchange interaction ($J_{\rm eff} \approx -0.006$ eV) places the system close to the instability between competing NM and AM states. The response to SOC is also strongly conditioned by the underlying exchange interactions. The defect-rich Feng *et al.* and Dai *et al.* samples exhibit a pronounced $\lambda$-independence of $T_N$, where $T_N$ is nearly insensitive to $\lambda$ in the Hiraishi et al. and Kiefer et al. samples. Moreover, no fully locked regime, defined by $T_N^{min}$ > 300 K together with detectable long-range magnetic order for all representative samples, is reached within the physically accessible SOC range ($\lambda \lesssim 0.2$) of the 4*d* oxide $RuO_2$. Extrapolation of the observed trend yields a critical value of $\lambda_c^{\rm (extrap)} \approx 22$, far beyond the physically accessible range, demonstrating that SOC alone cannot overcome the intrinsic sample-dependent magnetic competition. The magnetic phase selection therefore remains exchange-controlled throughout the physically accessible SOC regime.

Additional simulations (**Secs. 2.6** and **2.7** of the SM [36]) further show that increasing λ has little influence on the temperature dependence of $M_{\rm stag}$ or on the existence of long-range magnetic order. Instead, SOC primarily acts on the orientational degree of freedom of the ordered state, progressively locking the Néel vector to its preferred orientation and increasing the domain-wall energy, $E_{\rm DW} \propto \sqrt{|J_{\rm eff}|\lambda K_0}$, through the associated rotational anisotropy. Accordingly, the Feng *et al.* samples exhibit substantially larger domain-wall energies than the Hiraishi *et al.* samples, indicating stronger Néel-vector locking. These results establish a clear separation of roles between exchange interactions and SOC: exchange interaction primarily determines whether and which magnetic state emerges, whereas SOC primarily controls the orientation and stability of established magnetic order through Néel-vector locking and enhanced domain-wall energy. The disparate experimental observations can therefore be understood as different sample-dependent realizations of competing states within the exceptionally flat magnetic-energy landscape of $RuO_2$, rather than as contradictory observations of a unique magnetic phase.

Overall, our DFT calculations, machine-learning-assisted Hamiltonian construction, and large-scale MC simulations consistently reveal an exceptionally flat magnetic energy landscape in $RuO_2$, where the nonmagnetic state is nearly degenerate with multiple altermagnetic configurations. This near degeneracy originates from the delicate balance between localized exchange and itinerant magnetism, rendering the magnetic state intrinsically sensitive to perturbations of the local material environment. Intrinsic and extrinsic perturbations can therefore select distinct magnetic states primarily by modifying of the effective Heisenberg exchange interaction, while the perturbation-dependent Stoner channel provides an essential secondary contribution by renormalizing the relative stability of competing states. Although intrinsic defects are used here as a representative perturbation, the present framework can naturally be extended to other material perturbations, including lattice strain [30, 41, 42] and carrier doping [43, 44], which modify the local electronic environment and magnetic interactions. Such perturbations have also been explored experimentally in $RuO_2$ beyond the five representative sample conditions considered here. In contrast, spin-orbit coupling primarily governs the orientation and stability of the Néel vector rather than the emergence of long-range altermagnetic order. By providing a unified microscopic picture of the diverse magnetic behaviors observed in $RuO_2$, this work establishes a general microscopic framework for understanding how material perturbations select competing magnetic ground states on nearly flat magnetic energy landscapes, and provides quantitative guidance for the identifying and engineering altermagnetic materials.

*Conclusion*—In summary, we have shown that the long-standing controversy surrounding the magnetic ground state of rutile $RuO_2$ arises from an exceptionally flat magnetic energy landscape, in which the nonmagnetic state is nearly degenerate with multiple altermagnetic states. Within the effective spin-lattice framework developed here, material perturbations, including intrinsic defects, lattice strain, and carrier doping, select distinct magnetic states primarily by modifying the Heisenberg exchange interaction, while the perturbation-dependent Stoner channel provides an essential secondary contribution by renormalizing the relative stability of competing states. In contrast, within the physically accessible SOC regime, SOC primarily governs the orientation and stability of the Néel vector rather than the emergence of long-range magnetic order. These results provide a unified microscopic explanation for the diverse magnetic behaviors reported in $RuO_2$. More broadly, the framework developed here provides a general basis for understanding and controlling perturbation-induced selection among competing low-energy magnetic states in correlated materials, linking microscopic material environments to emergent magnetic phases.

*Acknowledgement*—Tianxiao Liang and Fanhan Kong contribute equal to this article. This work was supported by the National Natural Science Foundation of China (12274050 and 12534012) and Guangdong Provincial Quantum Science Strategic Initiative (GDZX2401002).

*Data available*—The data underlying this study is not publicly available, which can be accessed from the authors upon reasonable request.

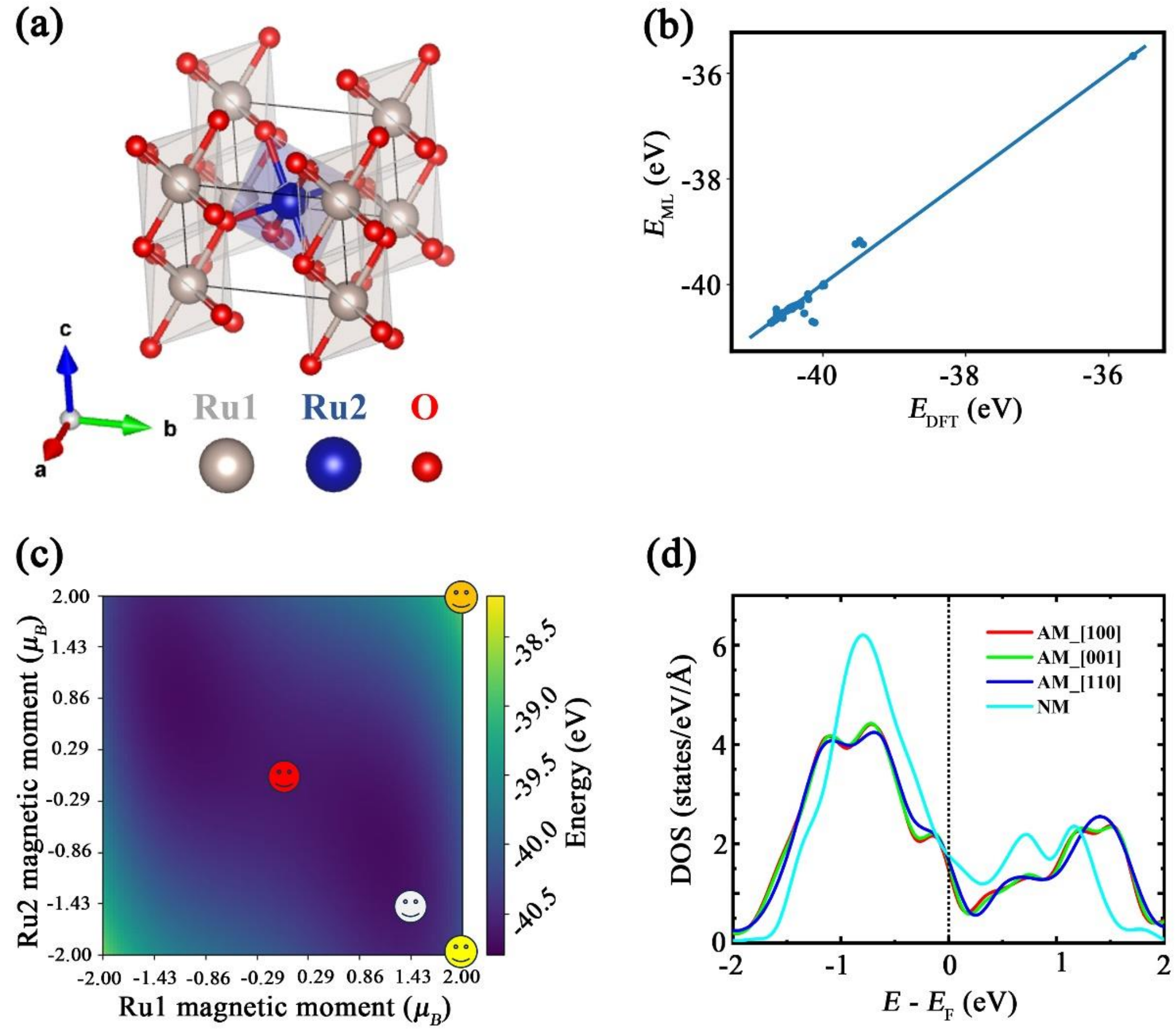

**FIG. 1. Flat magnetic energy landscape and competing magnetic states in rutile $RuO_2$**. (a) Crystal structure of rutile $RuO_2$ with two symmetry-related Ru sublattices, Ru1 and Ru2. (b) Comparison of DFT energies ($E_{DFT}$) with those predicted by the machine-learning-fitted generalized spin-lattice Hamiltonian ($E_{ML}$). (c) Magnetic energy landscape as a function of the signed local spin moments and on the two Ru sublattices. Positive and negative values correspond to opposite orientations of the local spin moment along $x$-axis. The red, yellow, white, and orange markers denote the NM, AM states with Néel vectors along [100] and [$1\bar{1}0$], and the FM state with magnetization along [100], respectively. The nearly degenerate NM, AM, and FM minima illustrate the exceptionally flat magnetic energy landscape of $RuO_2$. (d) Ru $d$-projected density of states (DOS) for representative magnetic configurations.

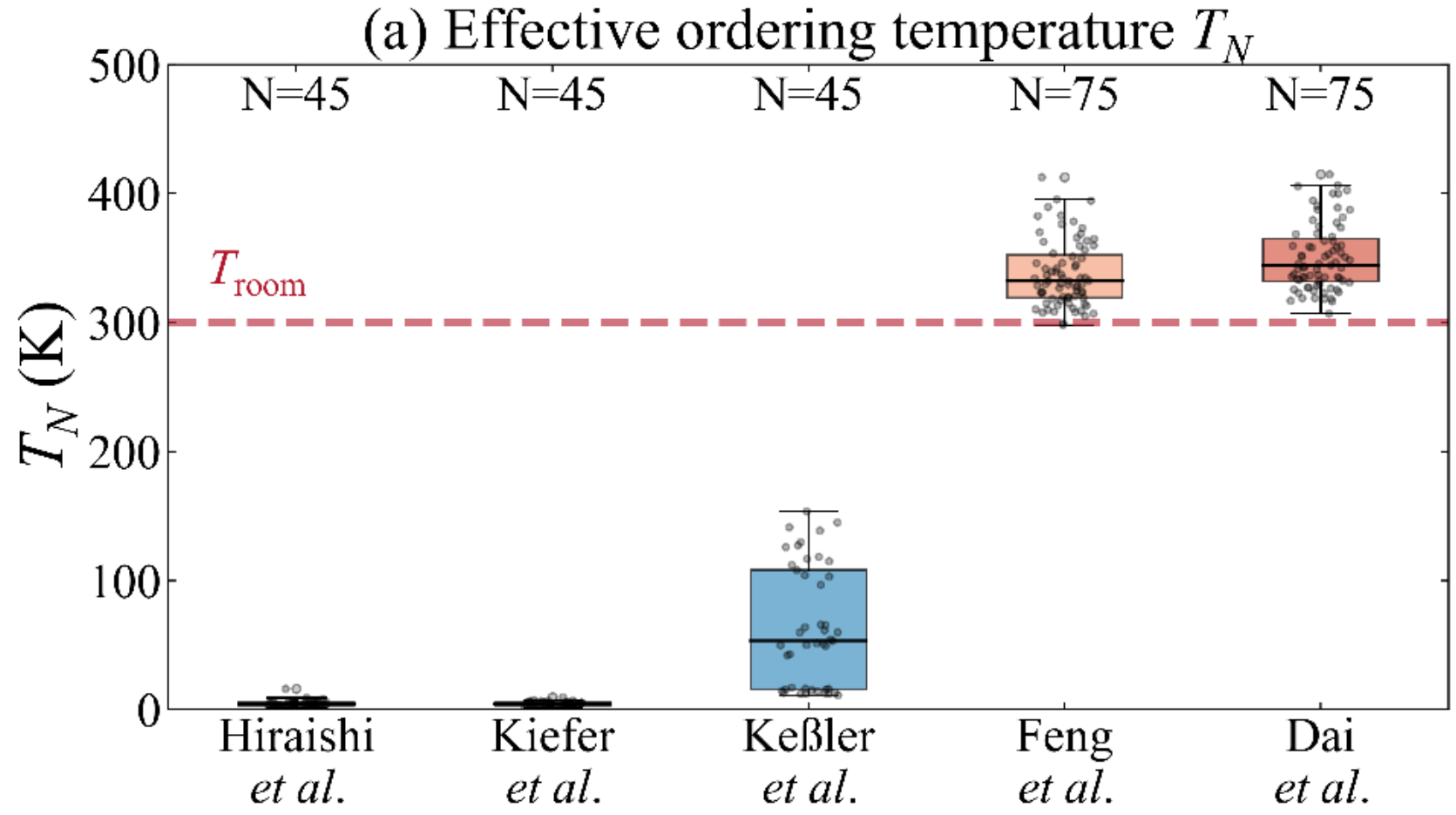

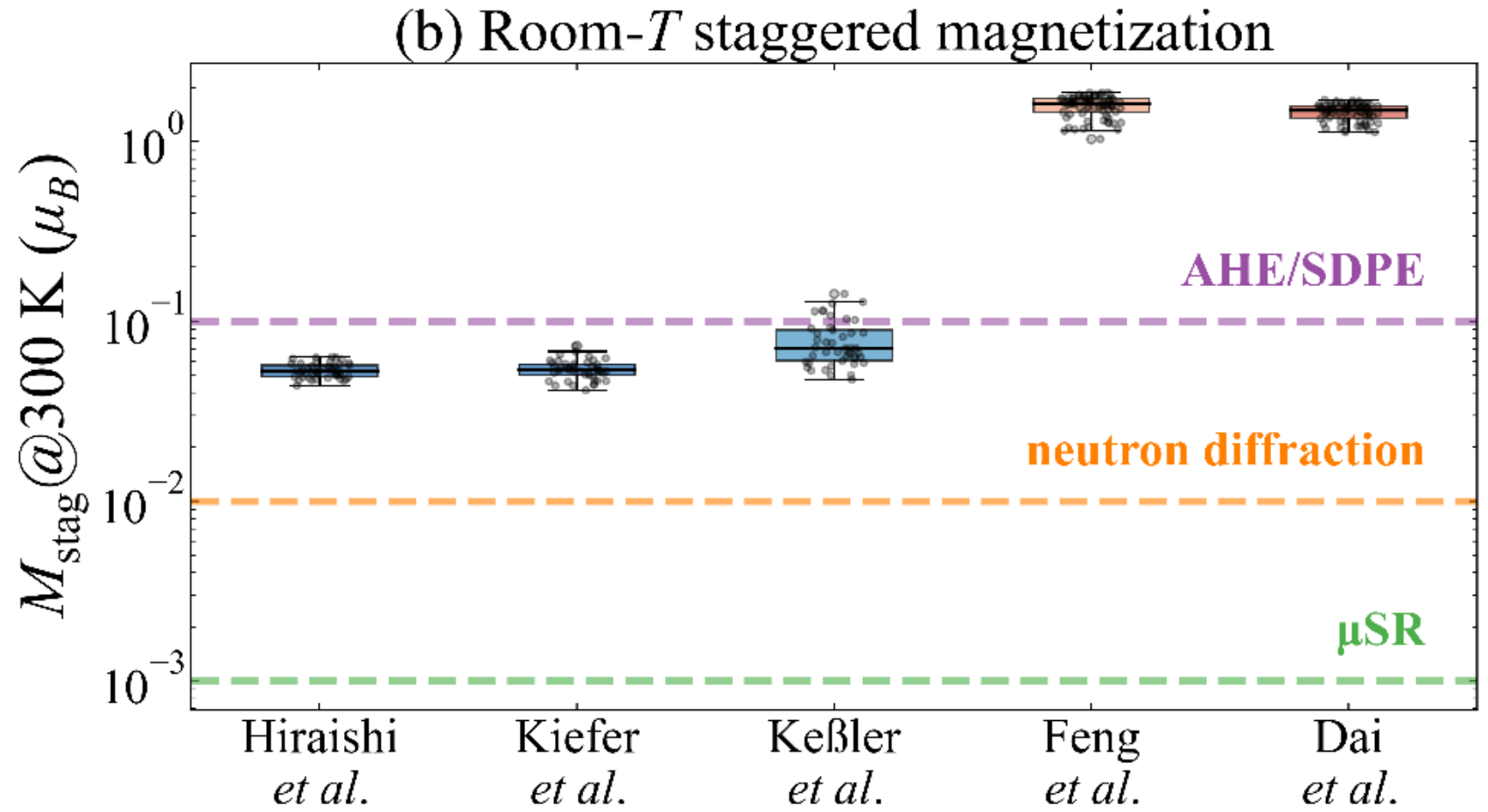

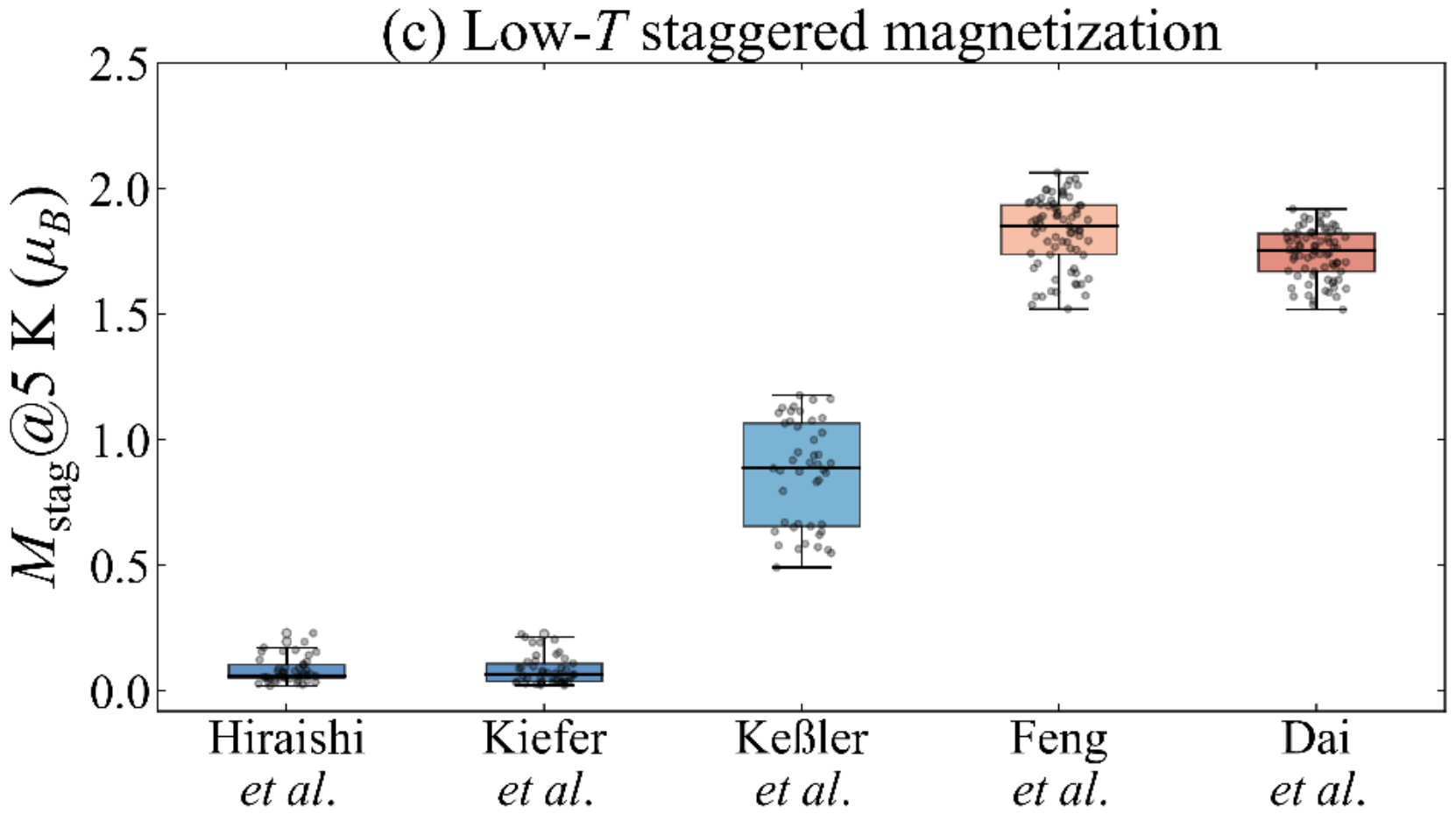

**FIG. 2. Sample-dependent magnetic order and experimental detectability in $RuO_2$.** (a) Effective ordering temperature $T_N$, extracted using the $T_{half}$ method described in the SM. N indicates the number of MC runs yielding a valid $T_N$ [36]); (b) Staggered magnetization at 300 K $M_{stag}$@300 K, show on a logarithmic scale; (c) Low temperature staggered magnetization $M_{stag}$@5 K. Horizontal dashed lines indicate the representative detection thresholds for AHE/SDPE ($10^{-1}$ $\mu_B$), neutron diffraction ($10^{-2}$ $\mu_B$), and $\mu$SR ($10^{-3}$ $\mu_B$).

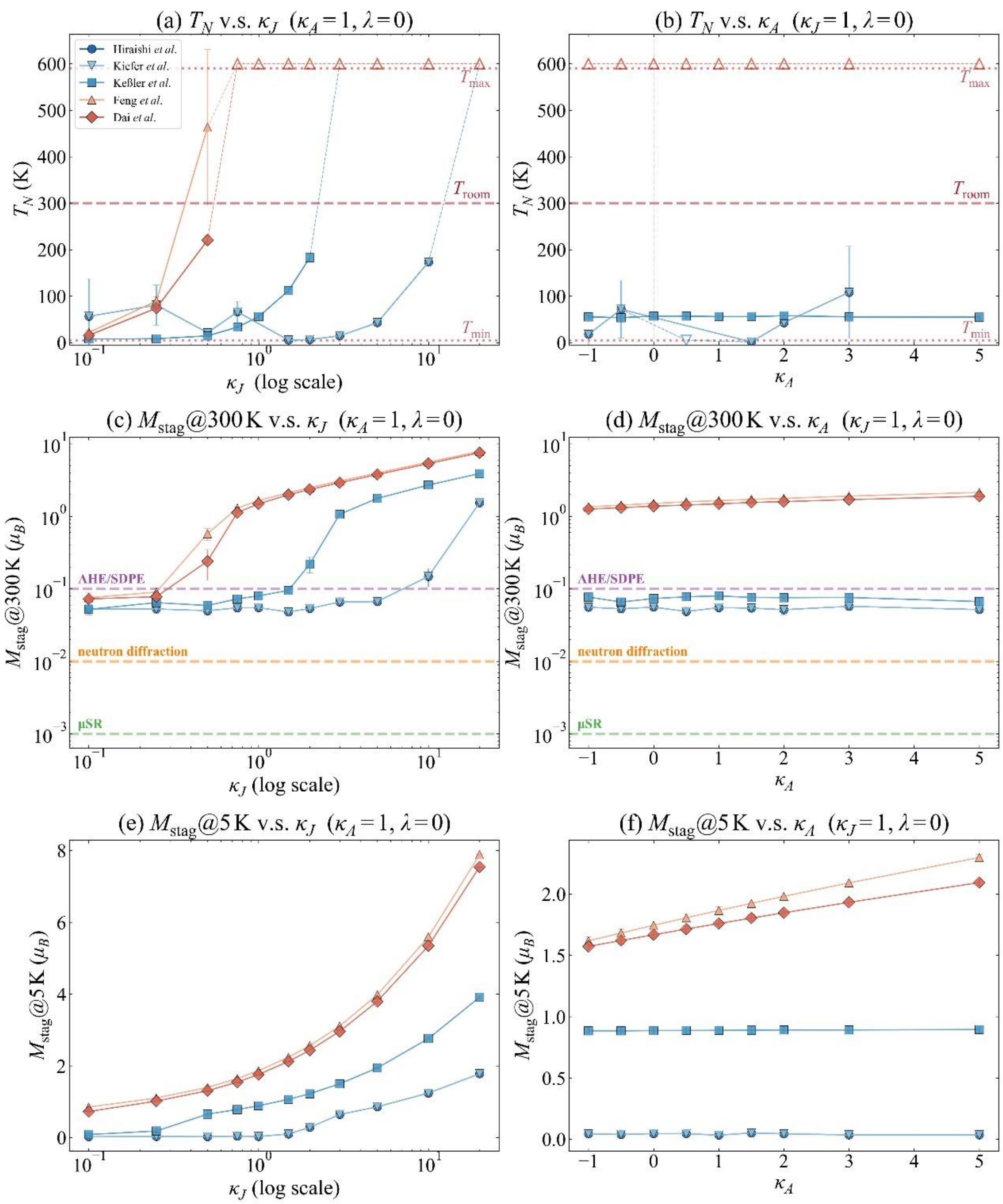

**FIG. 3. Independent modulation of the Heisenberg exchange and Landau channels in $RuO_2$ (λ =0).** (a) Effective ordering temperature $T_N$ as a function of the Heisenberg scaling factor $\kappa_J$ on a logarithmic scale for five representative experimental conditions at fixed $\kappa_A$ = 1. Values outside the simulated temperature window ($T_N$ > 590 K and $T_N$ < 5 K) are represented by proxy values and shown as open triangles connected by dashed lines. (b) $T_N$ as a function of the Landau scaling factor $\kappa_A$ at fixed $\kappa_J$ = 1. (c) Staggered magnetization at room temperature, $M_{stag}$@300 K, as a function of $\kappa_J$ on a logarithmic scale. (d) Low-temperature staggered magnetization, $M_{stag}$@300 K as a function of $\kappa_A$. (e) Low-temperature staggered magnetization, $M_{stag}$@5 K, as a function of on a logarithmic scale. (f) $M_{stag}$@5 K as a function of $\kappa_A$. Horizontal dashed lines in (c) and (d) denote the representative Horizontal dashed lines indicate experimental detection thresholds for AHE/SDPE, neutron diffraction and $\mu$SR.

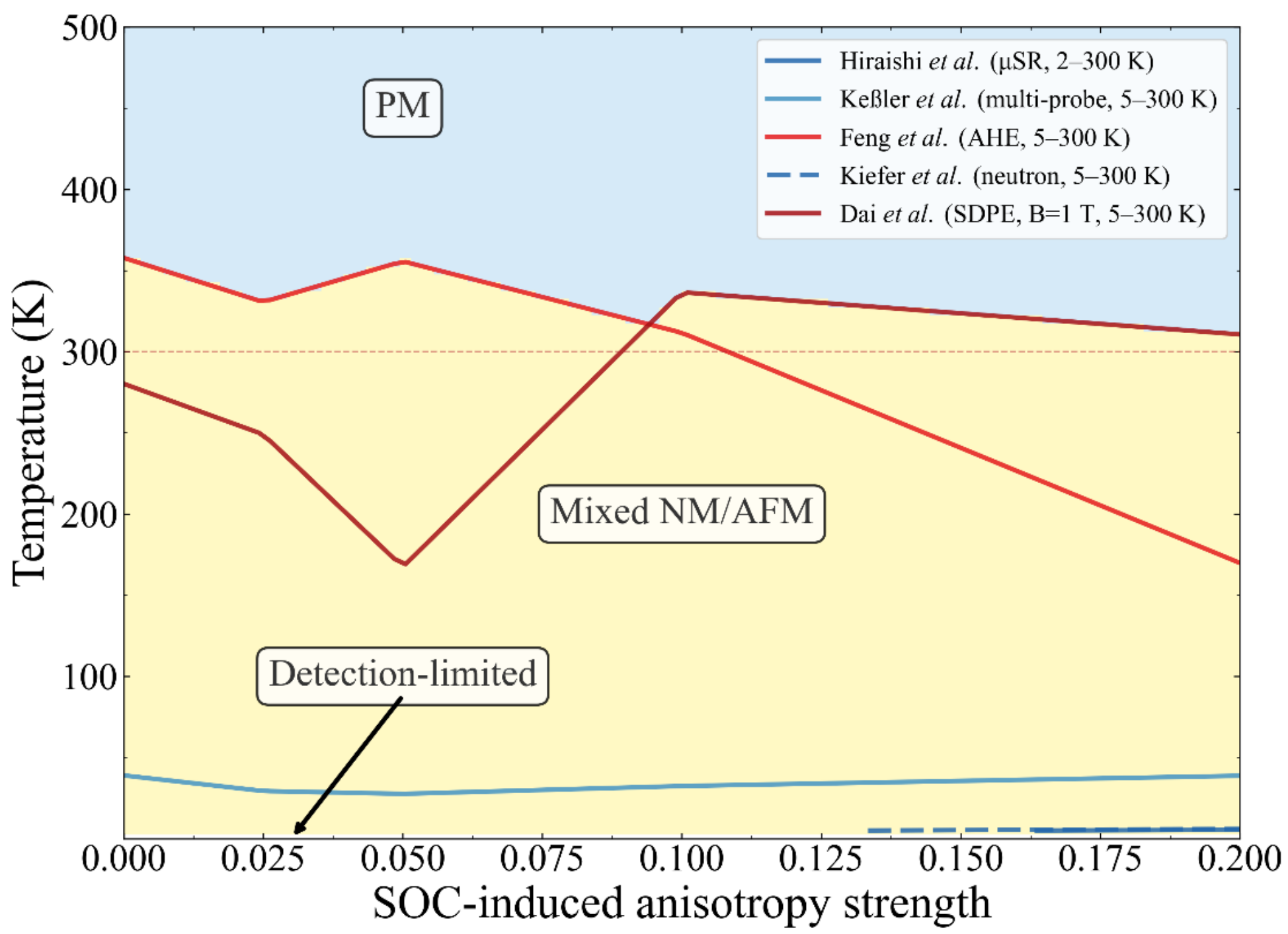

**FIG. 4. Temperature-SOC (*T*-*λ*) magnetic phase diagram.** Phase boundaries are determined by the calculated $T_N$-$\lambda$ curves for five representative experimental conditions [11, 20, 21, 23, 39]. The blue region corresponds to the paramagnetic (PM) phase for all samples, while the yellow region denotes the mixed magnetic regime, in which different experimental conditions stabilize distinct magnetic states despite nearly identical crystal structures. Its upper is defined by the $T_N$-$\lambda$ curves of the Feng/Dai *et al.* samples and is expected to extend beyond the simulated temperature range. The narrow low temperatures region ($\lambda \lesssim 0.1$ and $T \lesssim 5$ K) corresponds to a detection-limited regime, where magnetic order may exist but the ordered magnetic moment remains below the sensitivity of all currently available experimental probes. A fully locked regime, defined by $T_N^{\min} > 300$ K with robust long-rang magnetic order for all representative samples, is absent within the physically accessible SOC range. Linear extrapolation yields a critical value of $\lambda_c^{\text{extrap}} \approx 22$, far beyond the physically relevant regime for $RuO_2$.